\documentclass[aps,prd,floatfix,nofootinbib,showpacs,superscriptaddress,twocolumn]{revtex4-1}

\usepackage[utf8]{inputenc}
\usepackage{gensymb}
\usepackage{latexsym}
\usepackage{latexsym}
\usepackage{amssymb}
\usepackage{amsmath}
\usepackage{graphicx}
\usepackage{amsfonts}
\usepackage[mathscr,scaled=1.15]{urwchancal}
\DeclareFontFamily{OT1}{pzc}{}
\DeclareFontShape{OT1}{pzc}{m}{it}%
{<-> s * [1.15] pzcmi7t}{}
\DeclareMathAlphabet{\mathpzc}{OT1}{pzc}{m}{it}
\usepackage{supertabular}
\usepackage{placeins}
\usepackage{graphicx}
\usepackage{graphics}
\usepackage{bm}
\usepackage{color}
\usepackage{graphics}
\usepackage{epsfig}
\usepackage{float}
\usepackage{subfigure}
\usepackage{soul}
\definecolor{purple}{rgb}{0.5,0,0.5}
\definecolor{blue}{rgb}{0.0,0,0.9}
\usepackage[colorlinks=true, pdfstartview=FitV, linkcolor=purple, citecolor= purple, urlcolor=blue]{hyperref}

\begin{document}
	\title{Extracting a model quark propagator’s spectral density}
	\author{Zehao Zhu}
	\email{1710315@mail.nankai.edu.cn}
	\author{Kh\'epani Raya}
	\email{khepani@nankai.edu.cn}
	\author{Lei Chang}
	\email{leichang@nankai.edu.cn}
	\affiliation{School  of  Physics,  Nankai  University,  Tianjin  300071,  China }
	\date{\today}
	
	\begin{abstract}
		We propose a practical procedure to extrapolate the space-like quark propagator onto the complex plane, which follows the Schlessinger Point Method  and the spectral representation of the propagator. As a feasible example, we employ quark propagators for different flavors, obtained from the solutions of the corresponding Dyson-Schwinger equation (DSE). Two different truncations are employed. Thus, the analytical structure of the quark propagator is studied, capitalizing on the current-quark mass dependence of the observed features.
	\end{abstract}
	\maketitle 
	\section{Introduction}
	The strong-interactions part of the Standard Model, Quantum Chromodynamics (QCD), is characterized by two emergent phenomena: Dynamical Chiral Symmetry Breaking (DCSB) and confinement~\cite{Roberts:2016mhh,Roberts:2000hi}. DCSB is responsible for the vast majority of the mass of the visible universe and has a crucial impact on the observed hadron spectrum and properties; for example, it explains both the large mass of the proton and the unnaturally light mass of the pion. Confinement entails that colored states, such as QCD's fundamental degrees of freedom (quarks and gluons),  cannot appear in the spectrum. It also guarantees that condensates, typical order parameters of DCSB~\cite{Bashir:2008fk}, are wholly contained within hadrons~\cite{Brodsky:2012ku}. Thus, DCSB and confinement might be, in fact, intimately connected. Both phenomena can be potentially understood from QCD's 2-point functions~\cite{Roberts:2019ngp}, namely, propagators. Studying the analytical properties of the propagators could shed some light~\cite{Osterwalder:1974tc,Kugo:1979gm} on their confinement properties and the connection with DCSB and DSEs have been a cornerstone in handling such endeavors~\cite{Alkofer:2003jj,Maris:2002mt}. Thus, focusing on the matter sector, we obtain the quark propagator (in the space-like axis) through its corresponding Dyson-Schwinger equation (DSE)~\cite{Schwinger:1951ex,Schwinger:1951hq,Dyson:1949ha}, properly truncated and with interaction model inputs~\cite{Qin:2011dd}. Subsequently, the Schlessinger Point Method (SPM)~\cite{PhysRev.167.1411,Schlessinger:1966zz} is employed to extrapolate the propagator into the complex plane, allowing us to study the corresponding spectral function and its analytic structure. The article is organized as follows: in Section II we write the DSE for the quark propagator, the truncation and model inputs. Section III describes the SPM and the analytic continuation procedure. It is worth mentioning that the algorithm is quite general and can be employed to study different inputs, such as lattice QCD propagators. Section IV shows the numerical results and Section V summarizes the obtained conclusions.

	\section{Gap Equation}
	The DSE for the quark propagator, \emph{gap equation}, is the starting point for analyses of DCSB and confinement in the continuum, as well as the fundamental ingredient for hadron physics studies based upon Bethe-Salpeter or Faddeev equations~\cite{Maris:2003vk,Roberts:1994dr}. The gap equation, in Euclidean space, reads
	\begin{eqnarray}\nonumber
	S_f^{-1}(p)&=&Z_2(i \gamma \cdot p + m_f^{\textrm{bm}}) + \Sigma_f(p) \;,\\
	\Sigma_f(p) &=& \frac{4}{3}Z_1 \int_{dq}^{\Lambda} g^2 D_{\mu\nu}(p-q)  \gamma_\mu S_f(q) \Gamma_\nu^{f}(p,q)\;, 
	\label{eq:gapEq1}
	\end{eqnarray}
	where  $\int_{dq}^\Lambda = \int^\Lambda \frac{d^4q}{(2\pi)^4}$ stands  for  a  Poincar\'e  invariant  regularized  integration,  with  $\Lambda$  regularization scale. The rest of the pieces are defined as usual: $S_f$ is the $f$-flavor quark propagator, $D_{\mu\nu}$ is the gluon propagator and $\Gamma_\nu$ the fully-dressed quark-gluon vertex (QGV); $Z_{1,2}$ are the quark-gluon vertex and quark wave-function renormalization constants, respectively; $g$ is the Lagrangian coupling constant and $ m_f^{\textrm{bm}}$ is the bare-quark mass. The latter is related with the renormalization point ($\zeta$) dependent current-quark mass, $m_f^\zeta$, via Slavnov-Taylor identities~\cite{Slavnov:1972fg,Taylor:1971ff}. Each Green function involved obeys its own DSE, thus forming an infinite tower of coupled equations, which must be systematically truncated to extract the encoded physics~\cite{Binosi:2016rxz,Bender:1996bb}. Regardless of the truncation, a general solution for the fully-dressed quark propagator can be expressed as follows
	\begin{equation}
	\label{eq:quark-prop1}
	S_f(p) = Z_f(p^2)(i \gamma \cdot p + M_f(p^2))^{-1}\;,
	\end{equation}
	in analogy with its bare counterpart,
	$$S_f^{(0)}(p) =(i \gamma \cdot p + m_f^{\textrm{bm}})^{-1}\;.$$ 
	Here $Z(p^2)$ and $M(p^2)$ are dressing functions; the latter, independent of $\zeta$, is known as the mass function. In the Rainbow approximation~\cite{Qin:2011dd,Qin:2011xq,Maris:1999nt}, Eq.~(\ref{eq:gapEq1}) is modified according to the replacement:
	\begin{equation}
	\label{eq:Rainbow}
	g^2 Z_1 D_{\mu\nu}(p-q)\Gamma_\nu^f(p,q) \to Z_2^2 \tilde{D}_{\mu\nu}^f(p-q) \gamma_\nu\;, 
	\end{equation}
	where $ \tilde{D}_{\mu\nu}^f(k):=T_{\mu\nu}(k)\mathcal{G}^f(k^2)=(\delta_{\mu\nu}-k_\mu k_\nu / k^2) \mathcal{G}^f(k^2)$ and $ \mathcal{G}^f(k^2)$ is the effective coupling, expressed as~\cite{Qin:2011dd,Chen:2019otg}: 
	\begin{eqnarray}
	\label{eq:QinChang}
	\mathcal{G}^f(s=k^2) &=& \mathcal{G}_{\textrm{IR}}^f(s)+\mathcal{G}_{\textrm{UV}}^f(s)\;,\\
	\label{eq:QinChang2}
	\mathcal{G}_{\textrm{IR}}^f(s)&=& \frac{8\pi^2 D_f^2}{\omega_f^4} e^{-s/\omega_f^2}\;,\\
	\label{eq:QinChang3}
	\mathcal{G}_{\textrm{UV}}^f(s)&=& \frac{8\pi^2 \gamma_m \mathcal{F}(s)}{\ln [\tau + (1+s/\Lambda_{\textrm{QCD}})^2]}\;.
	\end{eqnarray}
	The term $\mathcal{G}_{\textrm{IR}}^f(s)$ provides an infrared enhancement, which is controlled by the product $\omega_f D_f^2$. Conversely, $\mathcal{G}_{\textrm{UV}}^f(s)$ is set to reproduce the one-loop renormalization-group behavior of QCD in the gap equation. We have defined: $s \mathcal{F}(s) = (1-\textrm{exp}[-s/(4m_t^4)])$, $\gamma_m = 12/(33-2N_f)$, $\tau = e^{10} - 1$, $m_t=0.5$ GeV, $\Lambda_{\textrm{QCD}}=0.234$ GeV and $N_f=5$. We solve Eq.~(\ref{eq:gapEq1}) for different current-quark masses, whose specific values and interaction strength are given in Table~\ref{tb:massparams}. 
	
	\begin{table}[htbp]
		\caption{\label{tb:massparams} One-loop evolved current quark masses and gluon model parameters. Masses, $\omega_f$ and $D_f$ are listed in GeV. The renormalization scale is set to $\zeta=2$ GeV.}
		\begin{center}
			\begin{tabular}{ccccc|cccccc}
				\hline \hline
				Flavor & $m_f^\zeta$ & $\omega_f$ & $D_f^2$ & &  & Flavor & $m_f^\zeta$ & $\omega_f$ & $D_f^2$ \\
				\hline
				$u/d$ & $0.005$ & $0.500$ & $1.060$ & & & $c$ & $1.170$ & $0.730$ & $0.599$ \\
				$s$ & $0.112$ & $0.530$ & $1.040$ & & & $b$ & $4.070$ & $0.766$ & $0.241$ \\
				\hline \hline
			\end{tabular}
		\end{center}
	\end{table}
	
	Key features of the gluon propagator, such as the infrared saturation and the connection with perturbation theory, are conveniently captured by the model defined in Eqs.~(\ref{eq:QinChang})-(\ref{eq:QinChang3}). However, we shall also explore the Renormalization-Group-Invariant (RGI) interaction described in~\cite{Cui:2019dwv}, which is derived in connection with the process-independent strong running-coupling, obtained from lattice QCD's Green functions at the physical pion mass. Thus, those attributes of the QCD coupling are captured from a first principles approach. In this case, the leading-order QGV, $\gamma_\nu$, turns out to be inadequate due to lack of tensor structures than enhance the contribution of the vertex to DCSB~\cite{Albino:2018ncl,Bashir:2011dp}. A minimal extension that makes the QGV compatible with~\cite{Cui:2019dwv} demands the inclusion of the Anomalous Chromomagnetic Moment (ACM) term,~\cite{Binosi:2016nme,Binosi:2016wcx,Chang:2010hb}.

	Thus, we explore a simple beyond-RL truncation (BYRL). In this case, the self-energy term in Eq.~\eqref{eq:gapEq1}, $\Sigma(p)$, is re-casted as follows ($k=p-q$):
	\begin{eqnarray}
	&&\Sigma(p)=Z_2 \frac{4}{3}\int_{dq}^\Lambda 4\pi \hat{d}(k)T_{\mu\nu}(k) \gamma_\mu S(q) \Gamma^{BY}_\nu(p,q)\;,\\
&&	\Gamma^{BY}_\nu(p,q) = Z_2\gamma_\nu + \eta \sigma_{\nu\alpha}k_\alpha \frac{B(p^2)-B(q^2)}{p^2-q^2} \mathcal{H}(k^2)\;,
	\end{eqnarray}
	where $k=p-q$, $B(p^2)=M(p^2)/Z(p^2)$; $(s/m_0^2)\mathcal{H}(s)=(1-e^{-s/m_0^2})$ is merely a profile function that controls the ultraviolet convergence and restricts the ACM effects to the infrared domain~\cite{Chang:2010hb}; $m_0= 2$  GeV and $\eta=0.37$ are parameters; and, finally, $\hat{d}(k)$ is the effective interaction from Ref.~\cite{Cui:2019dwv}. For the time being, in the case of the proposed BY truncation, we shall restricts ourselves to the $u/d$ quark.
	
	\section{Analytic continuation}
	\begin{figure}[t]
		\centering
		\includegraphics[width=0.7\columnwidth]{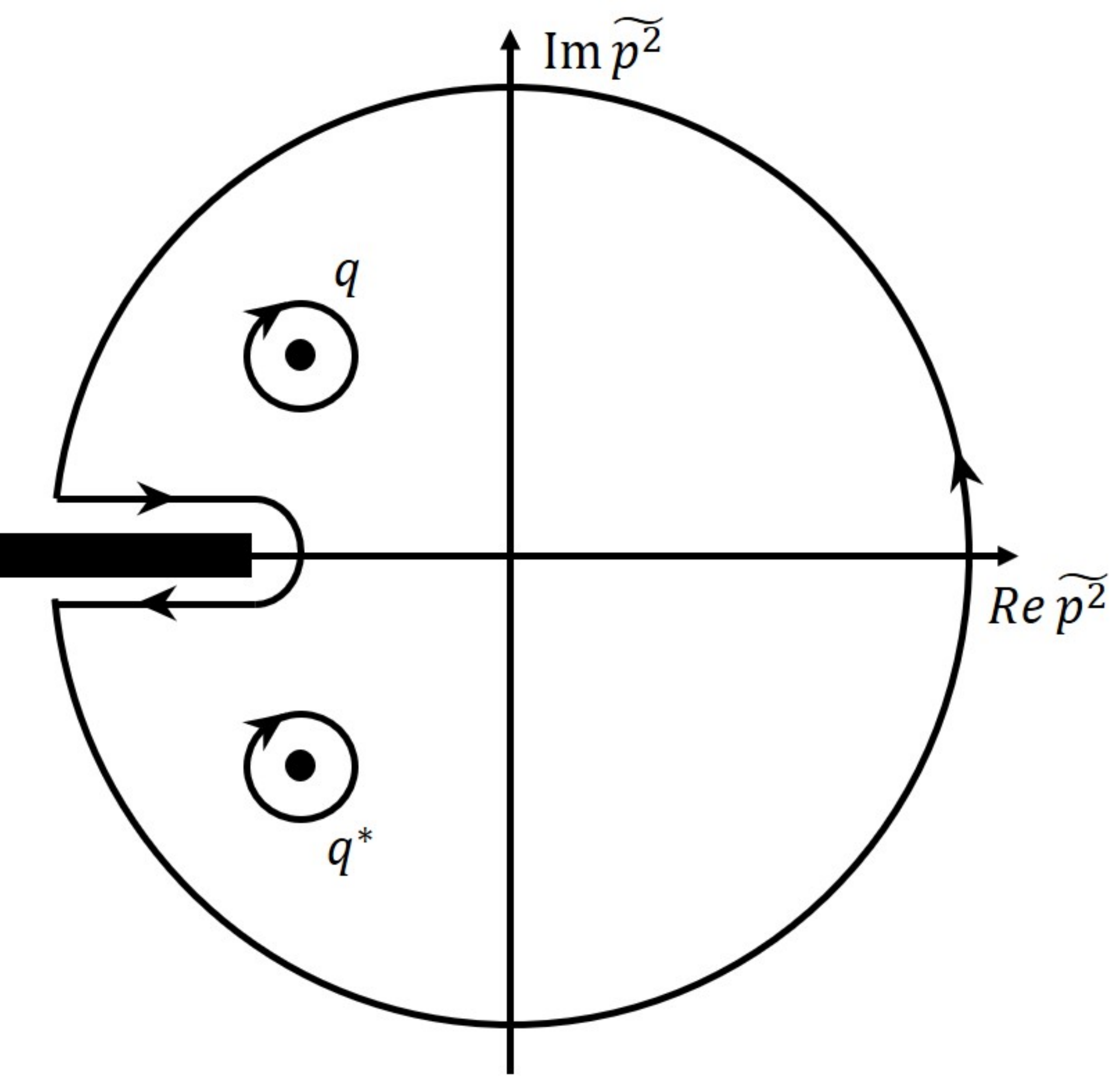}
		\caption{ The closed contour shown in this figure is used to calculate the Cauchy integral. This contour keeps all the poles outside and has a infinitesimal distance $\varepsilon$ away from the branch cut.}
		\label{Fig:contour}
	\end{figure}
	By solving the gap equation, it is \emph{almost} straightforward to obtain the quark propagator in the $p^2>0$ axis. The extension to the complex $p^2$-plane can be numerically challenging since, among other issues, one might encounter singularities~\cite{Jarecke:2002xd,Bhagwat:2002tx,dePaula:2017ikc,Chang:2019eob,Chen:2020ecu,Mojica:2017tvh}. Thus, we follow an analytic continuation scheme, based upon the spectral representation~\cite{Sauli:2004bx,Sauli:2006ba,Tripolt:2018qvi,Tripolt:2018xeo,Wang:2018osm} of the quark propagator and the SPM, to extrapolate our numerical solutions onto the complex plane.

	Firstly, it is convenient to re-express the quark propagator, in Eq.~(\ref{eq:quark-prop1}), in terms of its vector ($\sigma_V$) and scalar ($\sigma_S$) functions:
	\begin{equation}
	\label{eq:quark-prop2}
	S_f(p) = -i \gamma \cdot p \sigma_V(p^2) + \sigma_S(p^2)\;.
	\end{equation}
	Both $\sigma_{V,S}(p^2)$ are obtained on a large, discrete set of points ($p_i^2>0$, $i=1,\cdots,\; N$). Following the SPM, we employ a continued fraction representation such that 
	\begin{eqnarray}
	\nonumber
	\sigma(p^2)&=&\frac{\sigma(p_1^2)}{1^+}\frac{a_1(p^2-p_1^2)}{1^+}\cdots \frac{a_N(p^2-p_N ^2)}{1}\\ \label{eq:SPM1}
	&=& \sigma(p^2_1) \left[1+\frac{a_1(p^2-p^2_1)}{1+\frac{a_2(p^2-p^2_2)}{1+\cdots}} \right]^{-1},
	\end{eqnarray}
	interpolates (extrapolates) these functions (the labels $V,S$ are implicit). The coefficients $a_i$ are recursively obtained, ensuring that $\forall p_i^2$, the interpolated value of $\sigma(p_i^2)$ exactly reproduces the numerically obtained one,~\cite{PhysRev.167.1411}. Therefore, as discussed in Ref.~\cite{Tripolt:2018xeo}, the SPM gives almost exact reconstructions, of the continuous function, if the number  of  input  points  is  adequate. 
	
	Next we change the space-like $p^2$ to a complex value, $p^2 \to \tilde{p}^2 =  x + i y,\;(x,y\;\in \mathcal{R})$. An important feature of the SPM is its capability to identify singularities and branch points~\cite{Binosi:2019ecz,Tripolt:2017pzb}. However, it is seen that if $\tilde{p}^2$ remains real and positive in the training set, the SPM most likely will not introduce any singular structure besides scattered poles. This is due to the simple shapes of $\sigma_{v,s}$ on the space-like axis: finite, continuous and monotonically decreasing functions. Thus, it is found advantageous to deal with a  $\sqrt{p^2}$-grid instead~\cite{Binosi:2019ecz}. The training set is simply mapped as:
	\begin{equation}
	\label{eq:mapping}
	\{ p_i^2,\sigma(p_i^2)\} \to \{\sqrt{p^2_i},\sigma(p_i^2)\}\;.
	\end{equation}
	
	Clearly, this mapping introduces a branch point at $p^2=0$, and its corresponding branch cut along its negative real axis. Besides, other singular structures, in the form of complex conjugate poles (CCP), could also appear disperse on the complex plane~\cite{Jarecke:2002xd,Bhagwat:2002tx}. The closed contour, which leaves all the poles and branch cut outside, is sketched in Fig.~\ref{Fig:contour}. At this point, one can appeal to complex analysis theorems to represent $\sigma(p^2)$ in a convenient way. An arbitrary point inside the chosen contour can be written as a Cauchy integral. Moreover, from the residue theorem~\cite{Siringo:2016jrc,Siringo:2016vrv}, the quark propagator can be usefully re-expressed as follows:
	\begin{eqnarray} \nonumber
	\sigma(p^2)&=&\frac{1}{\pi}\;\int^{\infty}_{0}\frac{\rho(\omega)}{p^2+\omega}d\omega+\sum_i \left(\frac{R_i}{p^2-q_i}+ \frac{R_i^*}{p^2-q^{*}_i}\right)\;, \\
	\label{eq:Cauchy1}
	\rho(\omega)&=&\textrm{Im}[\sigma(-\omega-i\epsilon)]\;,
	\end{eqnarray}
	where $\epsilon$ is a positive infinitesimal real number; $R_i$ and $q_i$ take complex values, which are respectively identified as the residues and poles. The integral part is a standard dispersion relation for the propagator, such that $\rho(\omega)$ corresponds to its spectral function; the fraction part accounts for the possible presence of complex conjugate poles.

	In principle, $\rho(\omega)$ can be computed straightforwardly from the SPM on the numerical data. However, there are many components that can impact the stability of the extrapolations. We have identified key factors which alter the outcome: the number and distributions of points and, potentially, the mapping of Eq.~\eqref{eq:mapping}. Thus, in order to get a more accurate and stable result, we proceed as follows:
	\begin{itemize}
		\item Redistribute the big set of $N_p = m\cdot n$ points into $n$ subsets of $m$ points.
		\item Select one point from each subset, randomly, to form a new small set (of $n$ points).
		\item Implement SPM on the latter and obtain $\sigma(p^2)$.
	\end{itemize}
	This strategy makes it easy to control the number of points, while also covering most of the domain of the initial, much bigger, set. The SPM is performed several times, keeping the mean value as the final result and ensuring a small error is produced. Besides, the particular values of $m$ and $n$ are properly fixed by the requirement that the produced spectral functions are similar in shape and the location of the poles is stable. The latter is discussed below.
	\begin{figure}[t]
	\centering
	\begin{tabular}{c}
		\includegraphics[width=0.48\textwidth]{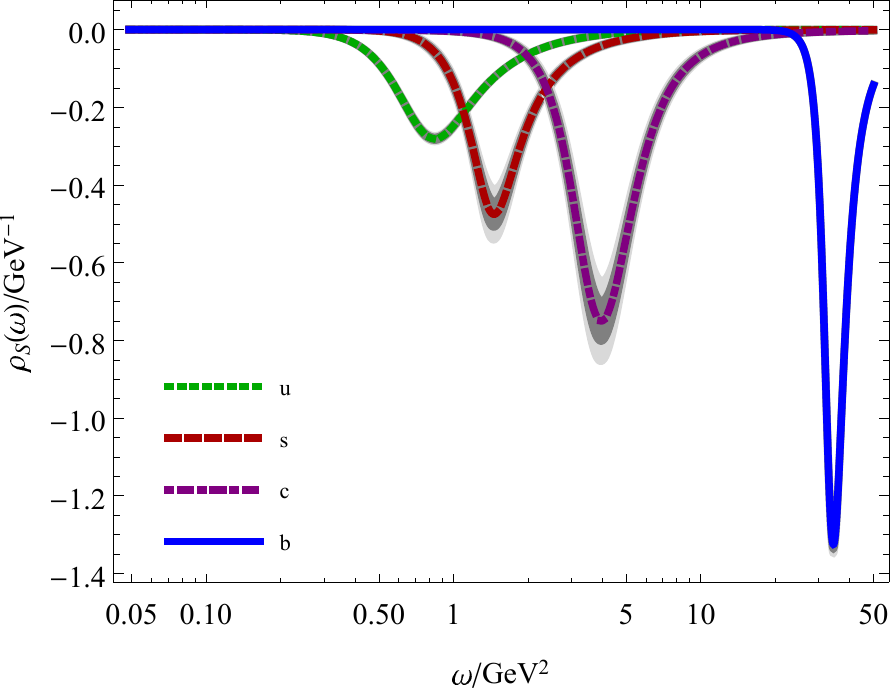}\\
		\includegraphics[width=0.48\textwidth]{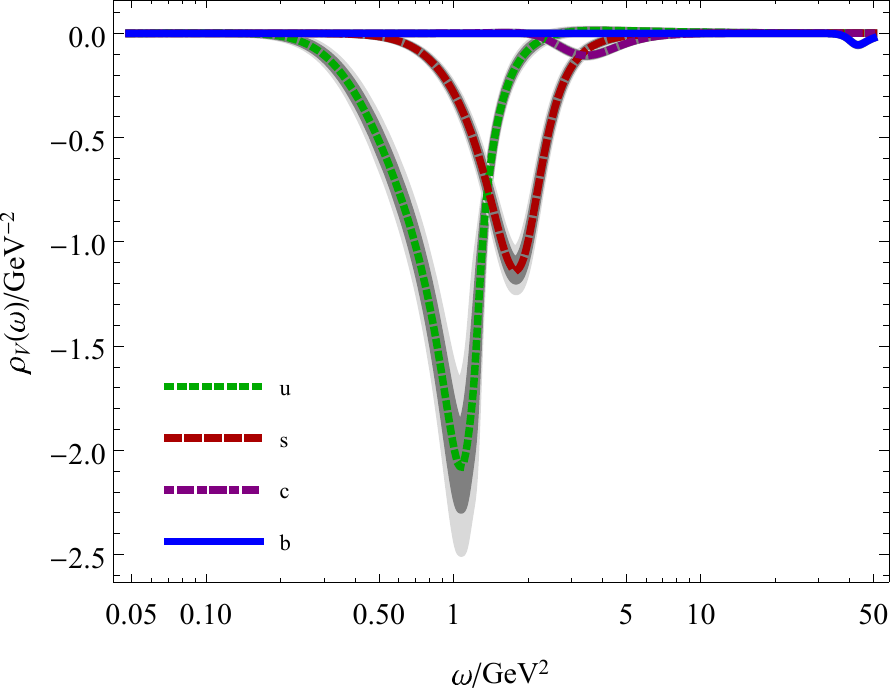}
	\end{tabular}
	\caption{[\textbf{Upper panel}] Spectral density associated with the scalar part of the propagator, $\sigma_s(p^2)$, in the RL truncation. [\textbf{Lower panel}] The analogous for the vector part. The dark-gray and light-gray shaded areas represent the $\sigma$ and $2\sigma$ confidence intervals, respectively. Consistently, the observed peaks are shifted towards $\omega\to \infty$ as the quark mass increases.}
	\label{Fig:rhosv}
	\centering
\end{figure}	
	
	Another important issue is the fact that $\rho(\omega)$ comes from the integral along the branch cut, but the branch cut might seem artificial, in the sense that it could appear only due to the choice of a $\sqrt{p^2}$-grid. This is a standard procedure~\cite{Binosi:2019ecz}, however. The presence of the branch cut serves a crucial purpose: it allows to write the propagator as in Eq.~\eqref{eq:Cauchy1}, enabling the access to its pole structure. An effective method to calculate the positions and residues of the poles has been already introduced in Refs.~\cite{Windisch:2016iud,Dorkin:2013rsa}, but it is not practical enough to reconstruct the propagator. For instance, we have seen that a single pair of CCP is sufficient to accurately determine the quark propagator. This implies that Eq.~\eqref{eq:Cauchy1} can be conveniently reduced to
	\begin{eqnarray} 
	\sigma(p^2)-\frac{1}{\pi}\;\int^{\infty}_{0}\frac{\rho(\omega)}{p^2+\omega}d\omega=\left(\frac{R}{p^2-q}+ \frac{R^*}{p^2-q^{*}}\right),
	\label{eq:Cauchy2}
	\end{eqnarray}
	such that, with $\rho(\omega)$ already determined, $q$ and $R$ (poles and residues) can be straightforwardly obtained following standard minimization procedures.

	\section{Numerical Results}
	From the solutions of the gap equation, Eq.~\eqref{eq:gapEq1}, for different quark flavors ($u/d$, $s$, $c$, $b$), one gets the space-like quark propagators. Then the corresponding spectral functions, poles and residues, are identified following the algorithm described in the previous section. The SPM is performed 50 times for each quark flavor; the outputs are averaged to produce a final result and error estimates. The particular values of $m$ and $n$, for each case, are specified in Table~\ref{tb:mandn}. 
	
	\begin{table}[htbp]
		\caption{\label{tb:mandn} Chosen $(m,n)$ values for each quark flavor; vector and scalar parts separately.}
		\begin{center}
			\begin{tabular}{cccc|cccc}
				\hline \hline
				Flavor & Vector & Scalar & & & Flavor & Vector & Scalar \\
				\hline
				$u/d$ & $(42,24)$ & $(16,60)$ & & & $c$ & $(28,36)$ & $(32,8)$ \\
				$s$ & $(10,24)$ & $(16,64)$ & & & $b$ & $(6,38)$ & $(42,24)$  \\
				\hline \hline
			\end{tabular}
		\end{center}
	\end{table}
	
\begin{figure}[t]
	\centering
	\begin{tabular}{c}
		\includegraphics[width=0.48\textwidth]{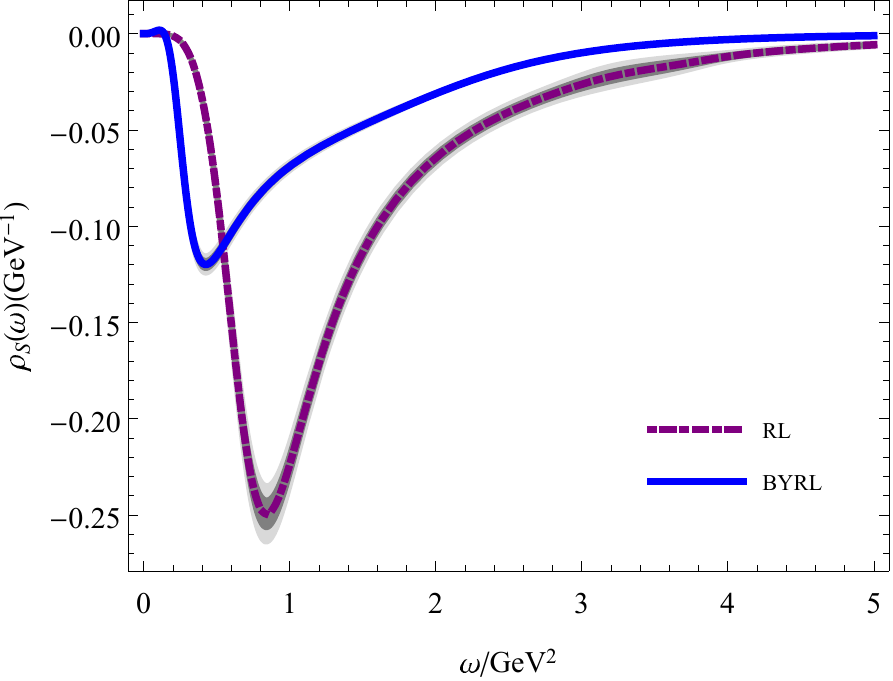}\\
		\includegraphics[width=0.48\textwidth]{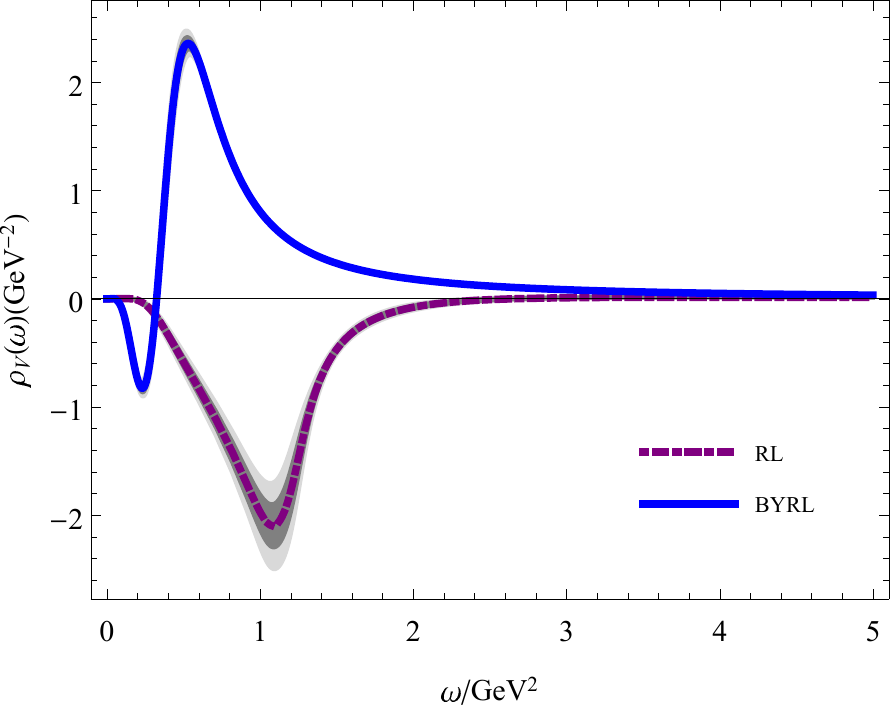}
	\end{tabular}
	\caption{[\textbf{Upper panel}] Comparison of RL and BYRL truncation results for the spectral density associated with the scalar part of the $u$-quark propagator, $\sigma_s(p^2)$. [\textbf{Lower panel}] The analogous for the vector part. The dark-gray and light-gray shaded areas represent the $\sigma$ and $2\sigma$ confidence intervals, respectively. }
	\label{Fig:rhosvBYRL}
	\centering
\end{figure}

\begin{figure}[t]
		\centering
		\includegraphics[width=0.48\textwidth]{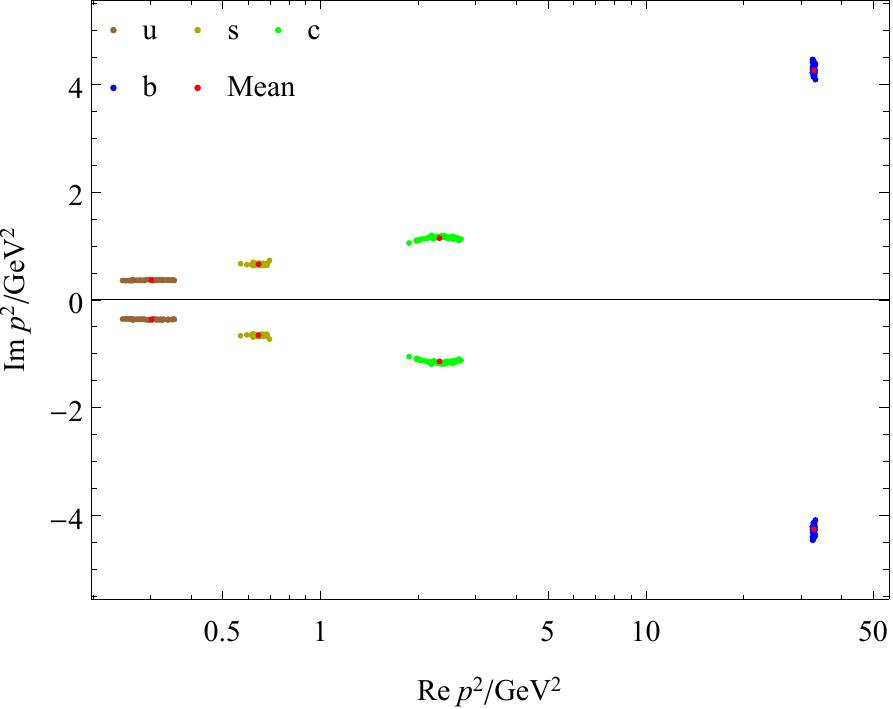}
		\caption{Position of the poles ($q$) in the complex plane, with extrapolation uncertainty. From left to right: $u$, $s$, $c$ and $b$ quark propagators. The displayed results correspond to the RL truncation.}
		\label{Fig:poles}
		\centering
	\end{figure}

\begin{figure}[t]
		\begin{tabular}{ccc}
			$u$-quark scalar part & $u$-quark vector part & $u$-quark propagator \\[2pt]
			\includegraphics[width=0.33\columnwidth]{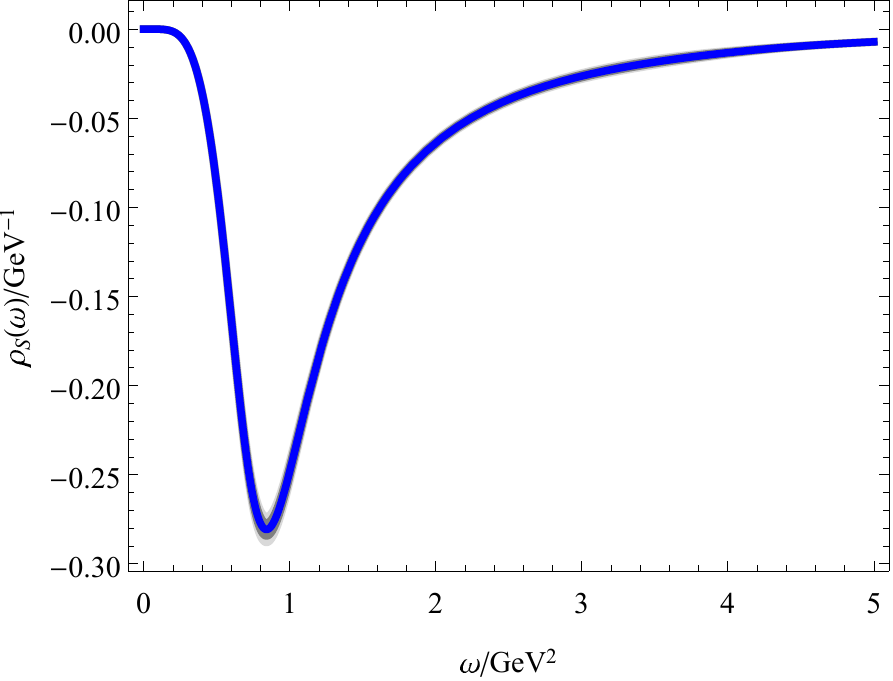} &   \includegraphics[width=0.33\columnwidth]{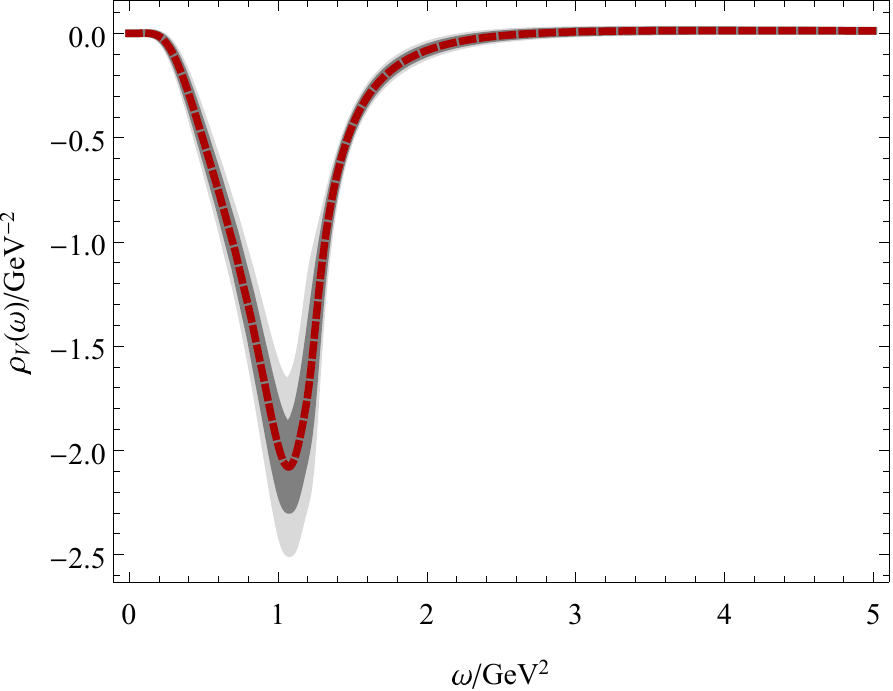} & \includegraphics[width=0.33\columnwidth]{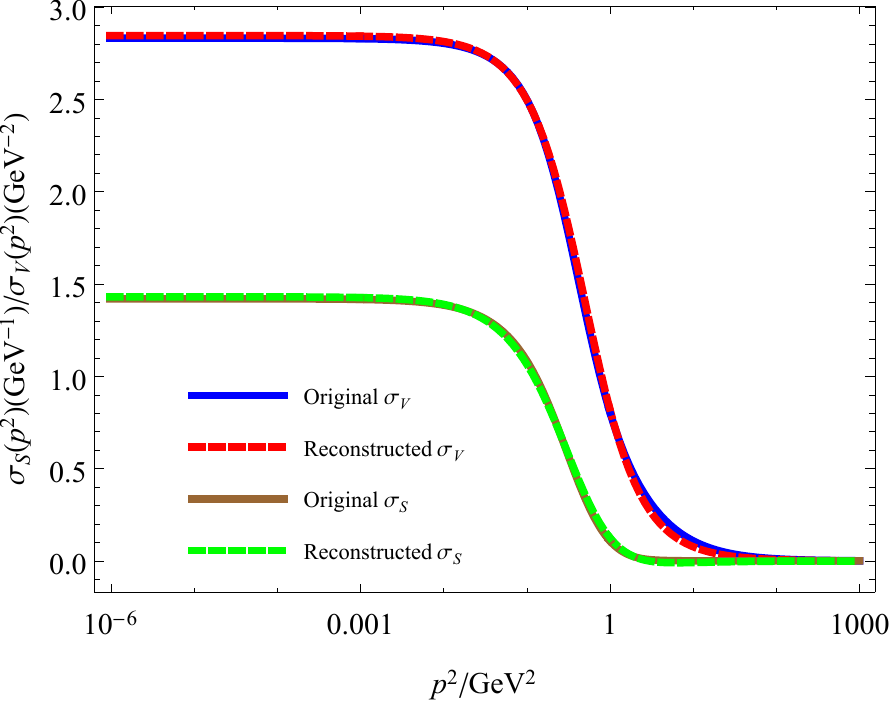} \\
			$s$-quark scalar part & $s$-quark vector part & $s$-quark propagator \\[2pt]
			\includegraphics[width=0.33\columnwidth]{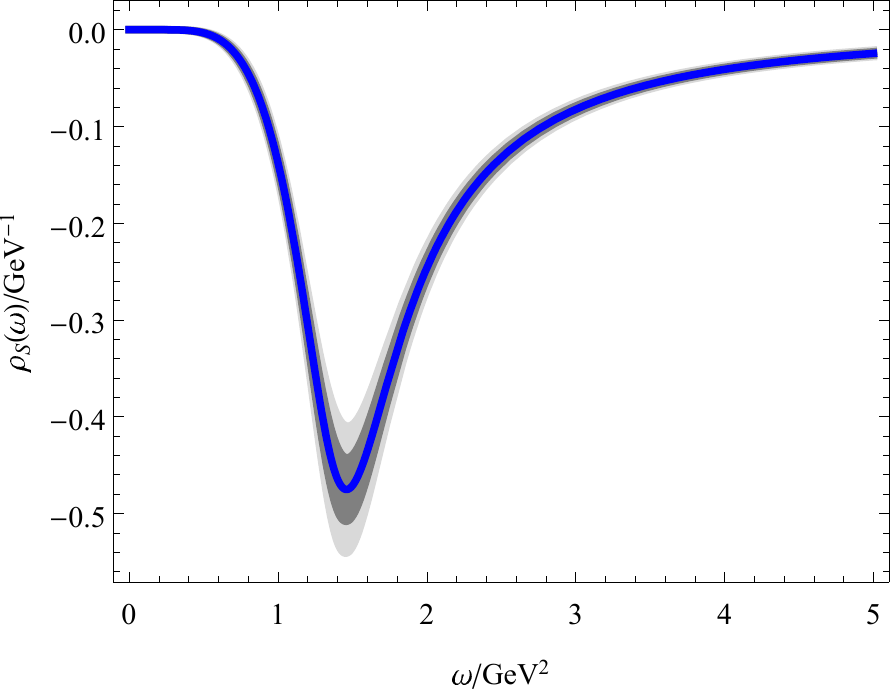} &   \includegraphics[width=0.33\columnwidth]{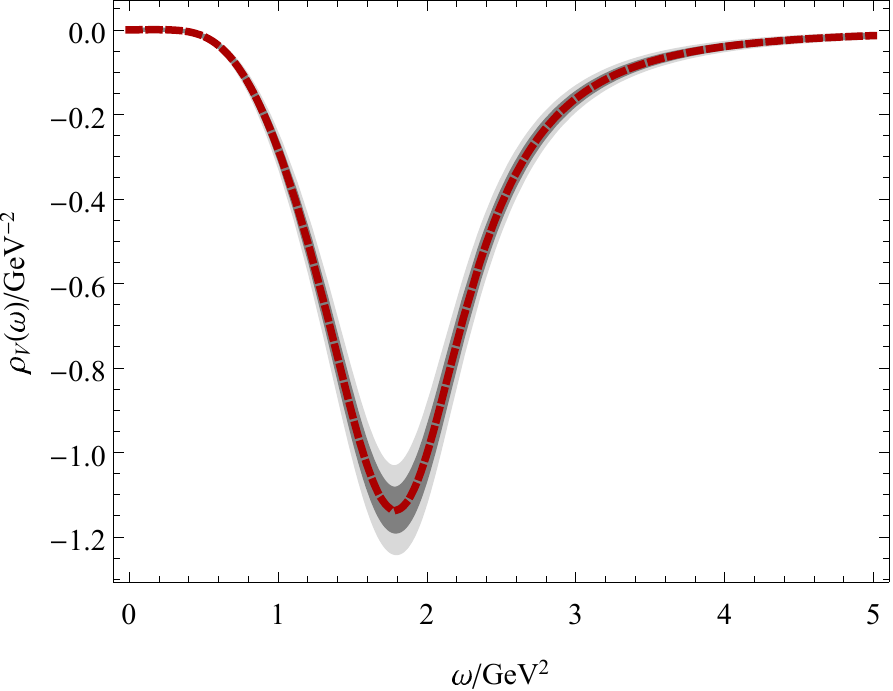} & \includegraphics[width=0.33\columnwidth]{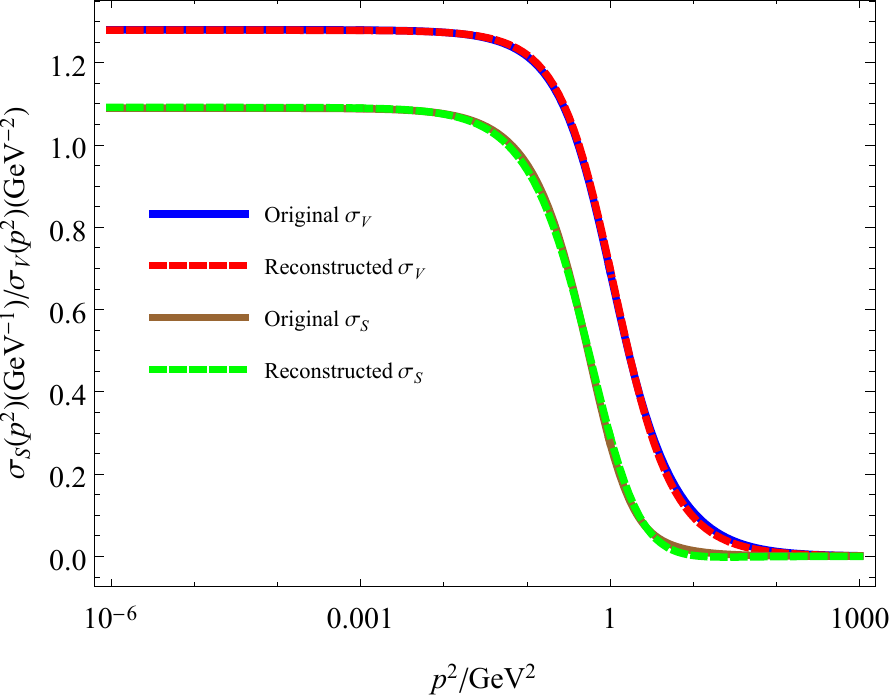} \\
			$c$-quark scalar part & $c$-quark vector part & $c$-quark propagator \\[2pt]
			\includegraphics[width=0.33\columnwidth]{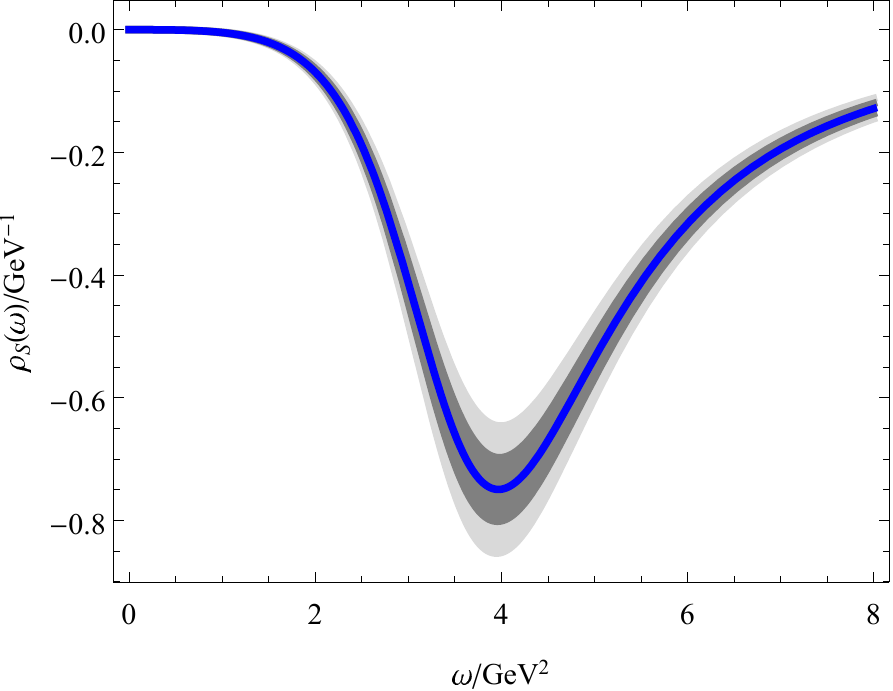} &   \includegraphics[width=0.33\columnwidth]{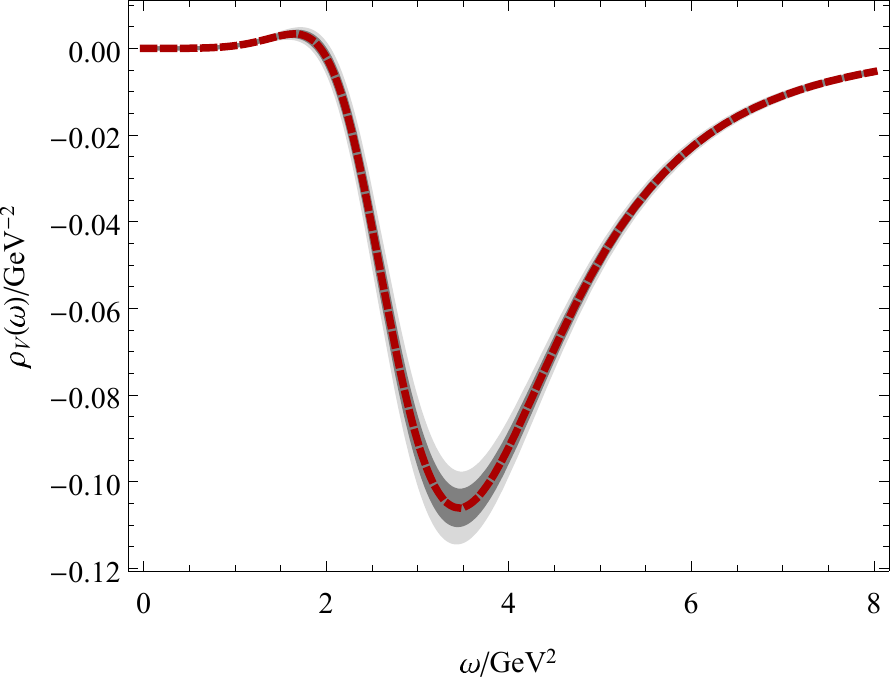} & \includegraphics[width=0.33\columnwidth]{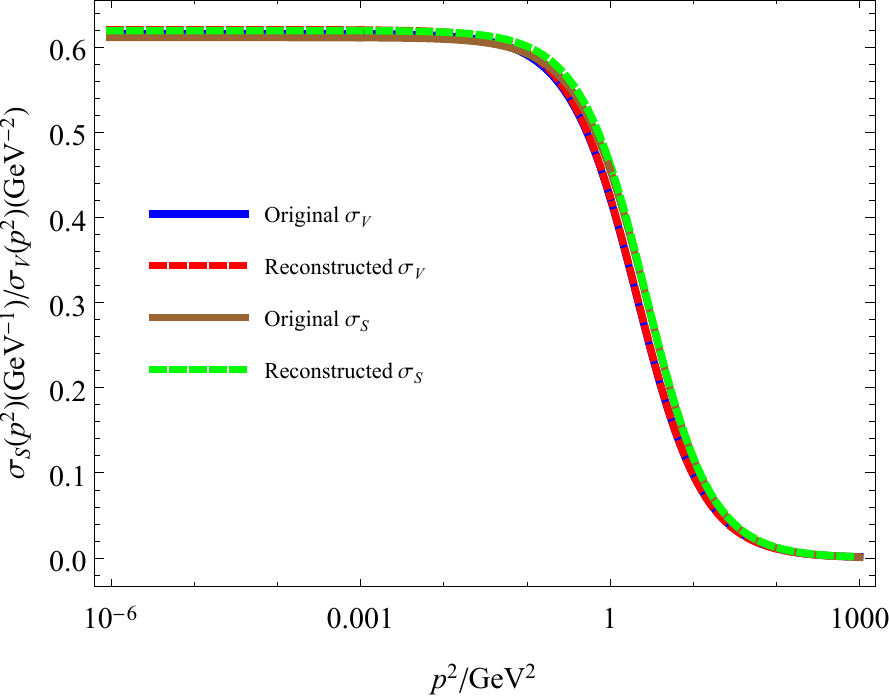} \\
			$b$-quark scalar part & $b$-quark vector part & $b$-quark propagator \\[2pt]
			\includegraphics[width=0.33\columnwidth]{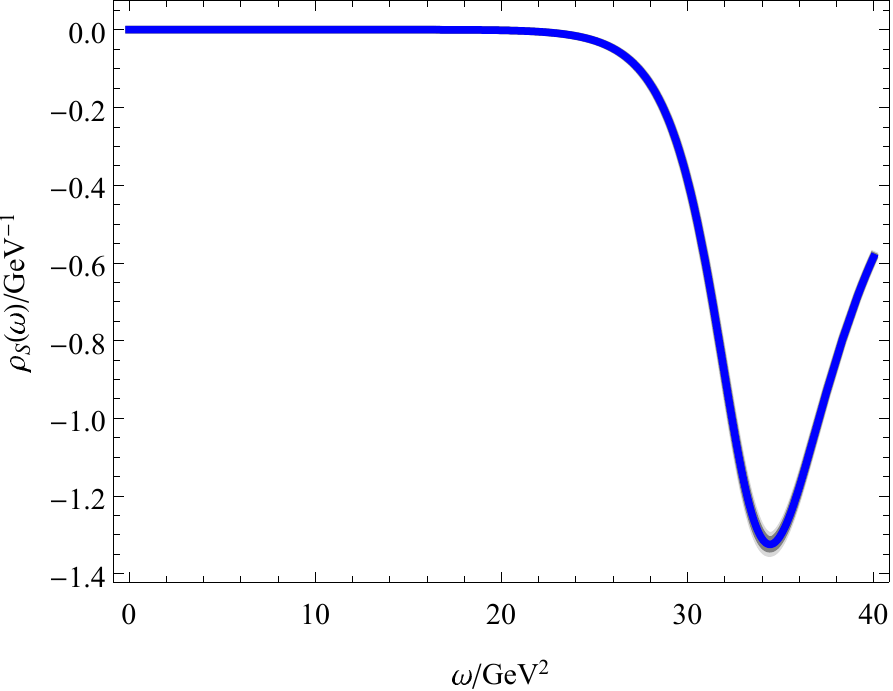} &   \includegraphics[width=0.33\columnwidth]{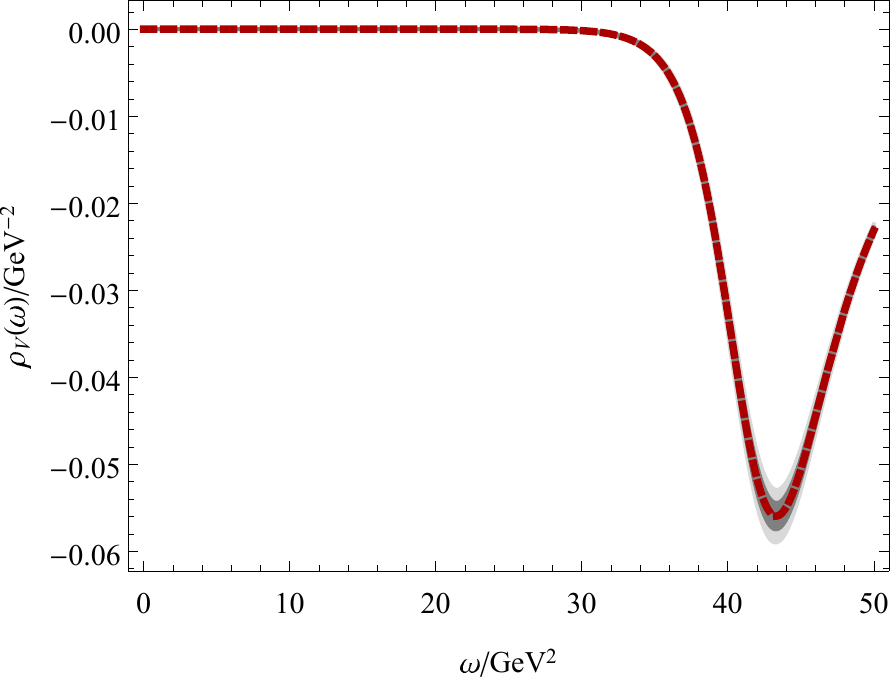} & \includegraphics[width=0.33\columnwidth]{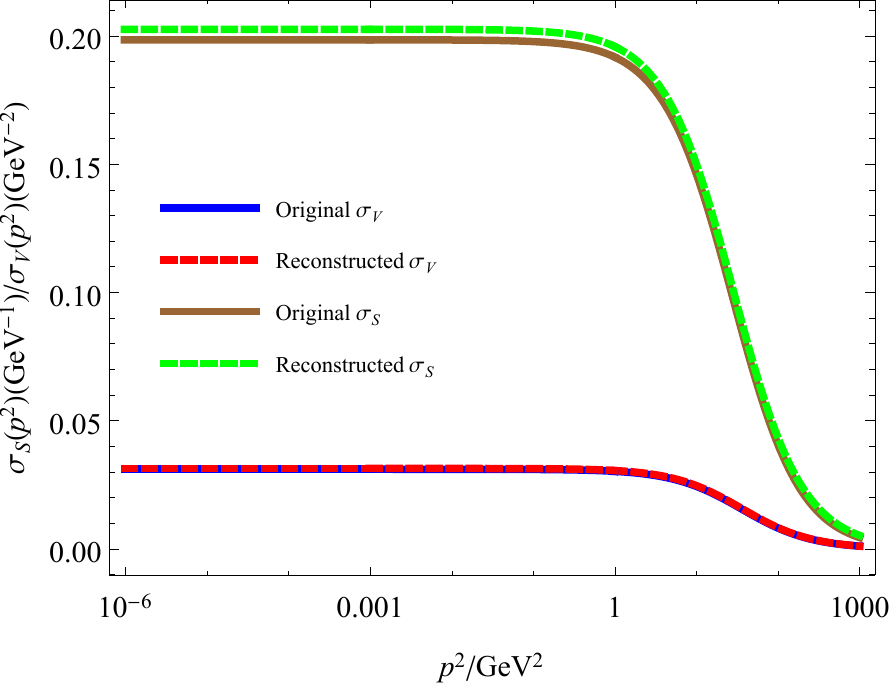} \\
		\end{tabular}
		\caption{The spectral densities and dressing functions for different quark masses, in the RL truncation. [\textbf{Left Panel}] Scalar part, $\rho_s(\omega)$. [\textbf{Middle Panel}] Vector part, $\rho_v(\omega)$. [\textbf{Right Panel}] Quark propagator dressing functions: $\sigma_s(p^2)$, $\sigma_v(p^2)$. From \textit{top} to \textit{bottom}: $u$, $s$, $c$ and $b$ quarks. The depicted spectral functions are the mean result after 20 times SPM and analytic continuation. The dark-gray and light-gray bands represent $\sigma$ and $2\sigma$ confidence intervals, respectively.}\label{Fig:props}
	\end{figure}

	The RL results for the spectral densities are displayed in Figs.~\ref{Fig:rhosv} and \ref{Fig:props}; the latter shows the curves separately, for each quark flavor, and also includes the quark propagator dressing functions. As it is clear from these figures, the spectral densities are not positive definite. Moreover, both $\rho_s(w)$ and $\rho_v(w)$ consistently exhibit a peak (a minimum).  Fig.~\ref{Fig:rhosvBYRL} compares the RL and BYRL results for the $u/d$-quark. Notably, $\rho_s(w)$ follow the same patter in both truncations and, although this does not happen for the $\rho_v(w)$, it is important to highlight that the produced spectral densities are not positive definite in either case. Focusing on the RL case, it is seen that the position of the peak moves towards $\omega\to\infty$ as the current quark mass increases. For the scalar part, the absolute value of this minimum increases with the quark mass, while the opposite pattern is observed for the vector part. A natural, kindred feature is also observed with the quark propagator dressing functions: at infrared momenta, the vector part of the $u$ quark is much larger that its scalar counterpart; this is completely reversed for masses above $m_{\textrm{cr}}\simeq m_c$. The position of the poles (with the associated uncertainty) as a function of the quark's mass is shown in Fig.~\ref{Fig:poles}. Their central values and the corresponding residues are captured in Table~\ref{tb:polesandres}. Clearly, the poles move deeper into the complex plain as the current mass grows, \emph{i.e.} it takes larger absolute values of both real and imaginary parts. This feature is consistent with realistic DSE studies~\cite{Chang:2013pq,Raya:2016yuj}.

\begin{figure}[h!]
		\centering
		\begin{tabular}{c}
			\includegraphics[width=0.48\textwidth]{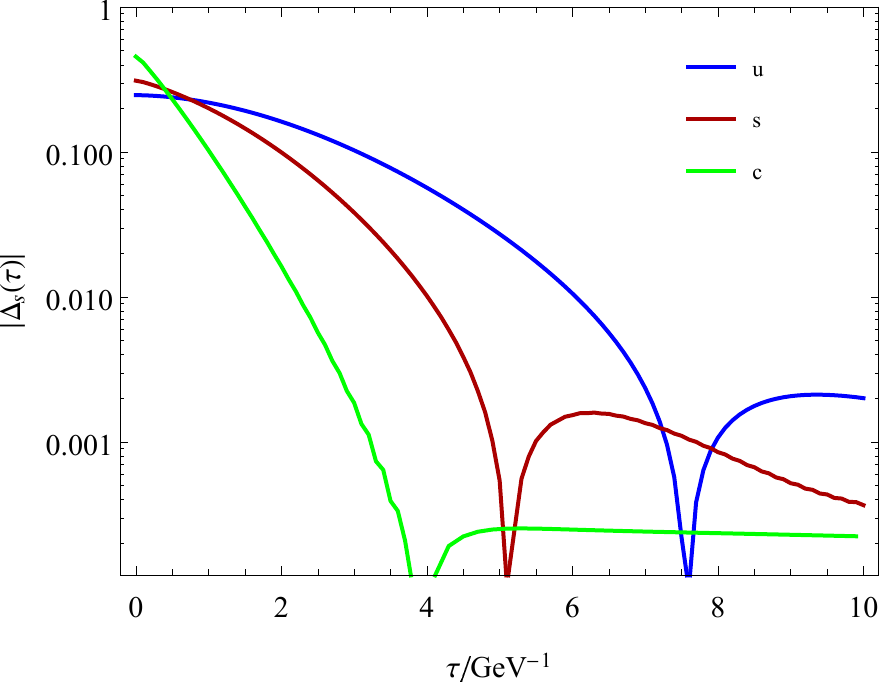}
		\end{tabular}
		\caption{[\textbf{Upper panel}] Space averaged Schwinger functions of the scalar part of the RL propagators: $u$, $s$ and $c$ quarks. The presence of the peaks, which moves towards $\tau \to 0$ as the quark mass increases, reveals a change of sign in $\Delta_s(\tau)$. The first peak of the $b$ quark propagator lies around $\tau \sim 1$, but the Schwinger function becomes numerically unstable for larger values of $\tau$.}
		\label{Fig:schwinger}
		\centering
	\end{figure}

	The non-positivity of the spectral functions (in particular $\rho_v(w)$) is often related to confinement~\cite{Osterwalder:1974tc,Alkofer:2003jj}. In this connection, we can also study the so-called spaced averaged (SA) Schwinger function~\cite{Bashir:2008fk,Bashir:2011vg}, which, at $\vec{p}=0$: 
	\begin{eqnarray}\nonumber
	\Delta(\tau)&:=&\int d^3x \int \frac{d^4p}{(2\pi)^4} e^{i(t p_4 + \vec{x}\cdot \vec{p})}\\
	&=& \frac{1}{\pi}\int_0^\infty dp_4 \cos (t p_4) \sigma(p_4^2)\;,
	\end{eqnarray}
	where $\sigma(p^2)$ is any of the scalar functions of the propagator. The SA Schwinger function for a real, $m$ massive, scalar particle will decay exponentially, $\Delta(\tau) \sim e^{-m \tau}$, since the propagator is simply $\sigma(p^2)=1/(p^2+m^2)$ and the mass shell can be reached in the real axis. On the other hand, if one has a propagator described by a pair of CCPs instead, one should expect an oscillatory behavior, $\Delta(\tau) \sim e^{-a \tau} \cos(b \;t + \delta)$ (here, $a$ is the real part of the CCP mass). In this case, the propagator could be associated with either a short-lived excitation that decays, or a confined, fundamental particle~\cite{Alkofer:2003jj}. Figure~\ref{Fig:schwinger} displays the SA Schwinger functions for the scalar parts of different quark flavors. The presence of the peaks in the logarithmic scale reveals a negative sign in the Schwinger function, thus another proof of violation of positivity. Naturally, the position of the peaks follow an opposite pattern, with respect to $\rho(\omega)$, \emph{i.e.}, peaks move towards $\tau \to 0$ with increasing quark mass.

	\begin{table}[htbp]
		\caption{\label{tb:polesandres} Poles ($q$) and residues ($R$) for different quark flavors. Top (bottom) panel corresponds to the RL (BYRL) truncation results. The mass units are expressed in appropriate powers of GeV.}
		\begin{center}
			\begin{tabular}{cccc}
				\hline \hline
				Flavor & $q$ & $R$ [$\sigma_v$] & $R$ [$\sigma_s$] \\
				\hline
				$u/d$ & $\;-0.302 \pm 0.364i\;$ & $\;0.586 \mp 0.542i\;$ & $\;-0.013  \mp 0.480i\;$ \\
				$s$ & $-0.646 \pm 0.660i$ & $0.702 \mp 0.311i$ & $0.060 \mp 0.719i$ \\
				$c$ & $-2.325 \pm 1.145i$ & $0.577 \mp 0.712i$ & $1.098 \mp 0.157i$ \\
				$b$ & $-32.942 \pm 4.260i$ & $0.674 \pm 0.498i$ & $5.110 \mp 3.287i$ \\
				\hline
				$u/d$ & $\;-0.175 \pm 0.210i\;$ & $\;0.231 \mp 0.685i\;$ & $\;0.001  \mp 0.390i\;$\\
				\hline \hline
			\end{tabular}
		\end{center}
	\end{table}

	\section{Conclusions and Summary}
	Starting with space-like quark propagators, we described a viable procedure to get access to their complex structure. The DSE inputs have been employed merely as an illustration and the proposed method is quite general. This is based upon the SPM, which interpolates the space-like quark propagator and extrapolates it into the complex plane. Then, a proper choice of integration contour allows us to rewrite the propagator in terms of a Cauchy integral, Eq.~\eqref{eq:Cauchy1}; thus defining the propagator's dressing functions in terms of spectral densities and granting us access to its pole structure. Remarkably, it is seen that a single pair of CCPs is sufficient to accurately represent the quark propagator. Among other things, this could expedite the computation of the form factors, which typically require two pairs of CCPs~\cite{Chang:2013nia,Raya:2015gva,Raya:2016yuj,Ding:2018xwy} to give a precise result. It is observed that the spectral densities are not positive definite and present peaks, which are shifted towards the ultraviolet region as the mass gets larger. Similarly, the position of the poles moves further into the complex plane with increasing current quark mass. The Schwinger functions exhibit the corresponding analogous features. Such consistency is encouraging. An immediate goal is to study whether or not the observed attributes  are still valid for truncations beyond the RL approximation. We have proposed a simple RL extension that makes use of the novel process-independent strong running-coupling~\cite{Cui:2019dwv} which, in contrast with the effective interaction that comes along with the RL truncation, is derived from a first principles approach to QCD's gauge sector. Our extension also includes the ACM term in the QGV, a crucial piece that is tightly connected with DCSB~\cite{Chang:2010hb,Bashir:2011dp}. Limiting ourselves to the $u/d$-quark, we have found that the produced spectral densities are not positive definite, as it also occurs with the RL truncation. This is a crucial characteristic that must prevail regardless of the current-quark mass. 
	
	Together with the practicality of our approach and the reduced error estimates, we believe that the algorithm discussed herein is also suitable to study other Green functions. Moreover, the SPM has been proven useful in connecting Euclidean and Minkowskian quantities~\cite{Ding:2019qlr,Ding:2019lwe}, a key goal in modern hadron physics~\cite{Liu:2019tkp,AlvarengaNogueira:2019zcs}, as well as a valuable extrapolation resource~\cite{Xu:2019ilh,Yao:2020vef,Cui:2020rmu}. The present approach to obtain the analytical representation of the quark propagator would take potential roles in the calculation of the hadron spectrum~\cite{Bhagwat:2002tx,Hilger:2017jti}, especially for the meson excited states~\cite{Chang:2019eob,Chen:2020ecu,Mojica:2017tvh}.

	\section{Acknowledgements}
	We are thankful to Daniele Binosi for sharing his code to implement the SPM. KR wants to acknowledge Alfredo Raya and Jos\'e Rodr\'iguez-Quintero for their valuable comments.
	
	\bibliographystyle{unsrt}
	\bibliography{bibliography}
	
	\clearpage
	\newpage
	\onecolumngrid
	
	\twocolumngrid

\end{document}